\documentclass[twocolumn,showpacs,preprintnumbers,amsmath,amssymb]{revtex4}

\usepackage{graphicx}
\usepackage{dcolumn}
\usepackage{bm}

\begin{document}


\title{Quasi-saddles as relevant points of the potential energy surface \\
in the dynamics of supercooled liquids}

%

\author{L. Angelani$^{1,2}$, R. Di Leonardo$^{1}$, G. Ruocco$^{1}$, A. Scala$^{1}$ and F. Sciortino$^{1,2}$}
\address{
         $^1$
Dipartimento di Fisica and INFM, Universit\`a di Roma ``La Sapienza'', P.le Aldo Moro 2, I-00185, Roma, Italy \\
         $^2$
{INFM - Center for Statistical Mechanics and Complexity, Universit\`a di Roma ``La Sapienza'', 
P.le Aldo Moro 2, I-00185, Roma, Italy.}
        }


\begin{abstract}
The supercooled dynamics of a Lennard-Jones model liquid is numerically investigated 
studying relevant points of the potential energy surface, i.e. 
the minima of the square gradient of total potential energy $V$.
The main findings are:
({\it i})  the number of negative curvatures $n$ of these sampled points 
appears to extrapolate to zero at the mode coupling critical temperature $T_c$;
({\it ii}) the temperature behavior of $n(T)$ has a close relationship with the 
temperature behavior of the diffusivity;
({\it iii}) the potential energy landscape shows an high regularity in the 
distances among the relevant points and in their energy location.
Finally we discuss a model of the landscape, 
previously introduced by Madan and Keyes [J. Chem. Phys. {\bf 98}, 3342 (1993)],
able to reproduce the previous findings.
\end{abstract}


\maketitle

\section{Introduction}
The supercooled liquid regime is the interesting state of the matter
which precedes the glass formation (once the crystal state is avoided).
Many efforts have been devoted to better understand the relevant 
physical processes taking place in the supercooled state,
the most important 
being the enormous increase of both relaxation times and inverse diffusivity
by many order of magnitude upon decreasing the temperature \cite{meta_liq}.
In recent years the numerical investigation of simple model liquids has provided
a very useful microscopic description of the supercooled regime and 
a detailed picture of the structural rearrangement of atoms during the dynamic evolution. 

One of the most powerful frameworks for the study of supercooled state
has been the so called Potential Energy Surface
(PES) description of the system \cite{deb_nature,wales_sci}.
This approach focuses on the properties of the multi-dimensional surface of the 
total potential energy sampled during the time evolution of the representative point in the 
$3N$-dimensional configuration space ($N$ is the total number of particles).
Different landscape features may control the motion of the particles 
\cite{angell,sastry,miller,keyes_cm,kur,thomas,lanave,heuer},
and the aim is to find those characteristics of the PES that have a direct 
relationship with (if possible the main causes of)
the emergent behavior of the relevant physical quantities.
The challenge is to find the ``{\it good}'' landscape features that 
one supposes to be responsible for the interesting phenomena, or 
at least useful to give a clear interpretation of them.
Among others, two different landscape approaches have been widely used in the last decades in this context:
the Instantaneous Normal Modes (INM) approach \cite{keyes}, based on the 
investigation of the PES very close to the instantaneous point in the configuration space 
during the molecular dynamics evolution of the system,
and the Inherent Structures (IS) approach \cite{stillinger}, based on the analysis of the minima of PES
visited by the system during its dynamic evolution.
In this paper we describe and revisit in a detailed way an additional approach, 
that has been very useful to give new insight in the analysis of the relevant 
processes taking place in the supercooled liquid regime. 
This approach focuses on the minima of the square gradient of the total potential energy,
``closest'' to the instantaneous points of the molecular dynamics trajectory.
This approach allows one to obtain a microscopic interpretation of 
the relevant processes, the main result being the characterization of the dynamics 
above and below the Mode Coupling Temperature $T_c$ \cite{mct}.
Moreover, the analysis of these points allows one to obtain information about some 
relevant characteristics of the PES, of great importance to construct  
simplified models of the landscape.

Lets us briefly resume the two main PES approaches, the INM and th IS approach. 
The INM method is based on the investigation of the PES around the instantaneous configurations 
${\bf r}$ (${\bf r}$ represents the $3N$-dimesional vector of the representative point in the configuration space)
during the molecular dynamics evolution of the system.
The diffusive quantities are supposed to be related to the shape of the energy surface at ${\bf r}$,
that is to say to the eigenvalues and eigenvectors of the Hessian matrix (the second derivative of the potential 
energy). 
The main hypothesis of the INM approach is that the relevant diffusive directions have to be searched 
among the downward curvatures (eigenvectors with negative eigenvalues).
Many attempts have been devoted to extend this approach to different liquid systems and 
to develop a theory of the supercooled liquid state based on INM concepts.
Moreover many efforts have been spent to recognize the true diffusive directions among all the downward ones,
as it is known there are downward curvatures that do not correspond to diffusive directions, 
notably in the crystalline state.
The INM negative curvature directions are classified as 
{\it shoulder modes} (related to anharmonicities of the PES) 
and {\it double wells} (with a double well shaped one-dimensional profile).
Diffusive directions are finally identified as those leading to different minima \cite{gezelter}
(this analysis involves the search of the minima of the PES).
The temperature dependence of the diffusion coefficient has been shown to be related to the temperature
dependence of the fraction of diffusive directions \cite{lanave_1}.
On the basis of this analogy, it has been conjectured a structural interpretation of the 
Mode Coupling Temperature $T_c$ as the temperature at which the number of 
diffusive directions goes to zero \cite{fs_DW}.

The second IS landscape approach is based on the analysis of the inherent structures, the minima of the PES.
To each instantaneous configuration ${\bf r}$ one associates a inherent structure ${\bf r}_{_{IS}}$
(using, for example, a steepest descent path starting from ${\bf r}$),
${\bf r} \rightarrow {\bf r}_{_{IS}}$,
partitioning the whole PES in the basins of attraction of minima.
This kind of analysis has been very useful not only in the study of the diffusive directions in the supercooled
regime (see the previous discussion about the INM), 
but also in the study of thermodynamic quantities 
(for example to evaluate the configurational entropy \cite{fs_entropy,colu}) 
and in the study of the out-of-equilibrium dynamics \cite{kob,fs_aging}.

In the present work we adopt a new approach in the analysis of the PES, in some way intermediate between the previous
two, in that it maintains information about the possible diffusive degrees of freedom (as the INM approach) 
and uses a mapping in the PES that associates instantaneous configurations ${\bf r}$ to new points
in the PES (as the IS approach).
In this approach \cite{noi_selle,cav_selle} the minima ${\bf r}_{_S}$
of the square gradient of the potential energy $W = |\nabla V|^2$
(saddles and some inflection points of the PES),
reached starting from equilibrated configurations ${\bf r}$, are calculated.
This procedure partitions the PES in basins of attraction of this new points:
${\bf r} \rightarrow {\bf r}_{_S}$.
The choice of $W$, suggested long time ago by Weber and Stillinger \cite{weber},
was motivated by the fact that saddles points of $V$ are absolute minima of $W$.
As pointed out recently by Doye and Wales \cite{doye_wales} and discussed in more details in the following, 
the numerical  minimization of $W$ locates mostly local minima of $W$, 
i.e. points which are not real saddles of $V$. 
For this reason in the following we refer to the local minima of
$W$ as Quasi Saddle Points (QSP) \cite{nota}.
In Ref.s \cite{noi_selle} and \cite{cav_selle} the topological and metrical properties of such points 
have been used for a description of the dynamics of supercooled liquids.
The investigated systems were monatomic Lennard-Jones and/or Binary Mixture Lennard-Jones. 
After equilibration of the system at a given temperature $T$, corresponding to
a given instantaneous potential energy $e$, 
the minima of $W$ were searched and associated to the stationary points of $V$ (saddles).
Then the energy $e_{_S}$ and the number of negative curvatures $n_{_S}$ at these saddle points were measured.
The dependence of $e_{_S}$ and $n_{_S}$ on the temperature \cite{noi_selle} or on $e$ \cite{cav_selle} show 
interesting properties:
({\it i}) the quantity $n_{_S}$ extrapolates to zero at $T_c$, demonstrating the validity of the conjecture 
that $T_c$ marks the transition between a dynamics among minima at low $T$ and a dynamics 
among saddles at high $T$;
({\it ii}) the aspects of the energy landscape seen by a given minimum is highly regular (as demonstrated 
by the linear dependence of $n_{_S}$ on $e_{_S}-e_{_{IS}}$ \cite{noi_selle} or on $e_{_S}$ \cite{cav_selle}).
The previous observation leads to the conclusion that the knowledge of the {\it saddle} points properties can be 
used to predict the supercooled dynamics and points out the relevance of these peculiar PES points.

Following these papers, the concept of {\it saddles} - as obtained by a minimization of $|\nabla V|^2$ -
has been further used in the investigation of the supercooled liquid dynamics \cite{shah,cav_nuovo}
and aging dynamics \cite{noi_aging}.

As we anticipated above, there is a drawback in the approach used in Ref.s \cite{noi_selle} and \cite{cav_selle}.
In fact, while it is true that in a saddle point (stationary point) of the $3N$-dimensional surface $V({\bf r})$,
the function $W$ has a minimum (actually, $W=0$ there and $W$ has a global minimum), the reverse is not guaranteed
to be true. There are local minima of $W$, with $W \neq 0$, that are NOT stationary points of $V$.
In these points, along one or more directions ${\bf r} (\xi)$, parametrized by $\xi$,
the function $V(\xi)$ has an inflection point.
In these inflection points (QSP), the Hessian of $V$ has a number of vanishing
eigenvalues (excluding the three translation directions)
equal to the number of directions ${\bf r} (\xi)$ where $V$ has an inflection point.
This property has been used in \cite{noi_selle} in order to identify the QSP:
a threshold was arbitrary chosen, and all the minimizations of $W$ that lead to points where the Hessian of $V$ 
has at least one eigenvalue smaller (in modulus) than the threshold were discarded from the analysis.
In \cite{noi_selle} we claimed that only a small fraction of minimizations ended in a QSP points, 
and we estimated in  $2 \%$ the error introduced in $n_{_S}$ by the choice of the threshold.

Doye and Wales \cite{doye_wales} have shown - by the analysis of the same system investigated in 
\cite{noi_selle} and \cite{cav_selle}, the BMLJ with periodic boundary conditions - 
that almost all (around $95 \%$ in their work)
the minimizations of $W$ ended in a QSP point, and not in a true saddle as claimed in \cite{noi_selle} and 
\cite{cav_selle}.
Motivated by such contradictory results, we reanalyzed the minima of $W$ found in \cite{noi_selle}.
In this work we show the result of such an analysis.
We conclude that the results of Doye and Wales are correct and 
the error in the analysis in \cite{noi_selle} originates from the ``imperfect'' minimization of $W$:
in a point close to (but not exactly at) QSP, the Hessian of $V$ no longer has ``small'' eigenvalues,
and the criterion used to identify false saddles fails.

One puzzling question, however, still remains open: why the quantity $n_{_S}$ - i.e. the number of negative
eigenvalues at the minima of $W$ - is related to the critical temperature of the system ?
It remains true that $n_{_S}(T)$ extrapolates to zero at $T_c$.
In the present paper we address this question, and we will further stress
the importance of the quantity $n_{_S}$;
we will show that also the transport properties (namely the diffusion coefficient) can be
determined by the knowledge of $n_{_S}$.

The paper is organized as follow:
In section II we describe the new approach, defining the saddles of the PES 
and studying very carefully their operative definition and the problems related to 
the presence of false saddles.
Then in section III we expose the relevant results obtained with the new PES approach, 
giving a topological interpretation of the mode coupling dynamic transition.
In section IV the relationship between diffusivity and saddle properties is evidenced,
giving a simple interpretation of the diffusion processes in terms of saddle order.
In section V we describe some landscape features as emerged from saddle analysis.
In section VI a simplified model of the PES is described, together with its relevant properties
compared to the features of model liquids investigated above.
At the end (section VII) a brief summary of the main results and conclusions are reported.

\section{Quasi Saddle Points}

The analyzed system is a simple model liquid, a monatomic Modified Lennard-Jones 6-12 (MLJ). 
The model is able to support strong supercooling without the occurrence of crystallization, due to the presence
of a small perturbation term in the Hamiltonian that inhibits ordering (this term is
a function of the static structure factor, see Ref. \cite{noi_fdt} for more details).
The system is composed of $N=256$ particles enclosed in a cubic box with periodic boundary conditions.
Truncated ($R_c=2.6$) and shifted LJ potential are used.
Equilibrium configuration are prepared by standard microcanonical molecular dynamics simulations at constant density 
$\rho = 1$ (standard LJ units are used hereafter) and at temperatures ranging from the normal liquid phase 
($T \sim 1.6$) down to the mode coupling temperature ($T_c = 0.475$, estimated from diffusivity).

\begin{figure}[htb]
\includegraphics[width=.45\textwidth,angle=0]{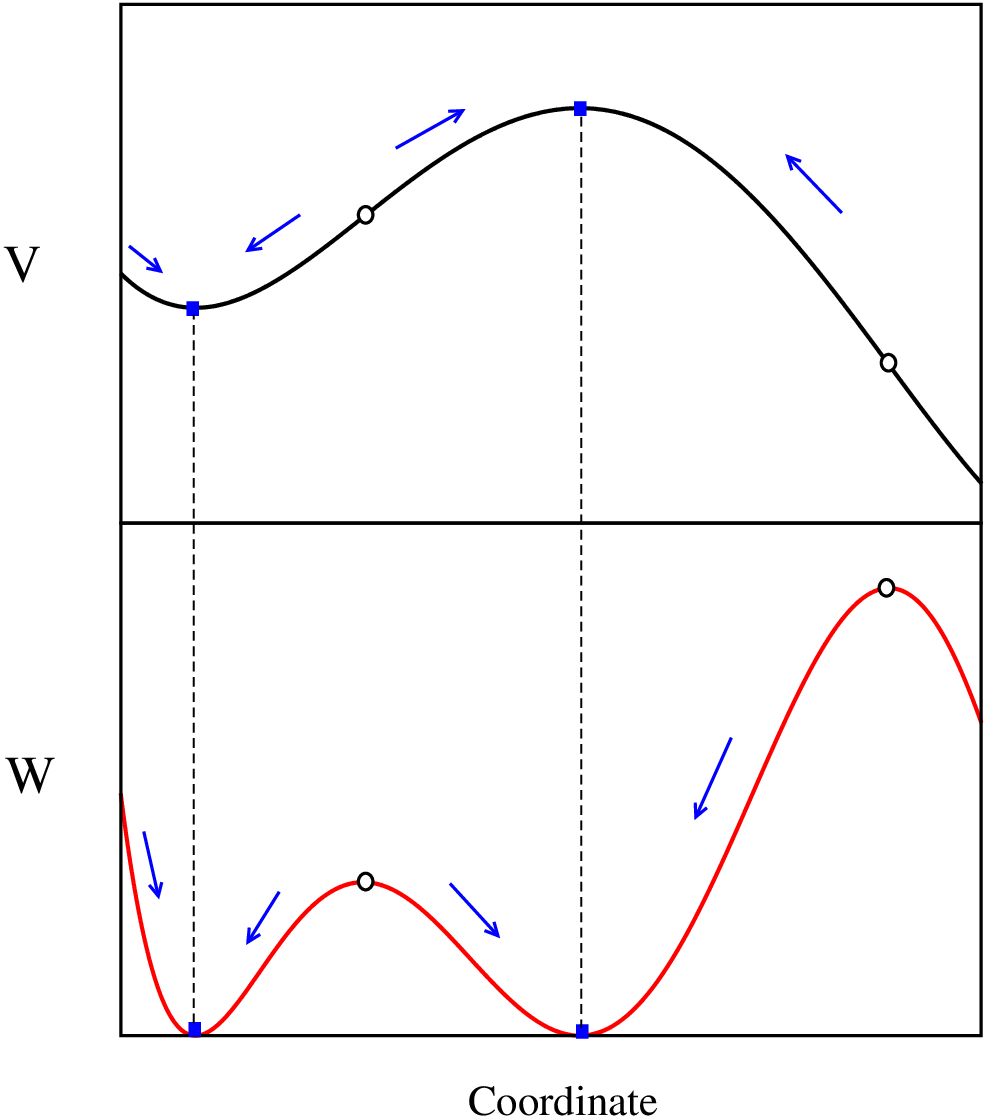}
\caption{Example of profile of $V$ and $W$ along a given direction. The minimum and maximum of $V$ correspond
to absolute minima of $W$. The arrows indicate the basins of attraction of the minima of $W$.}
\label{fig_pro_1}
\end{figure}

\begin{figure}[htb]
\includegraphics[width=.486\textwidth,angle=0]{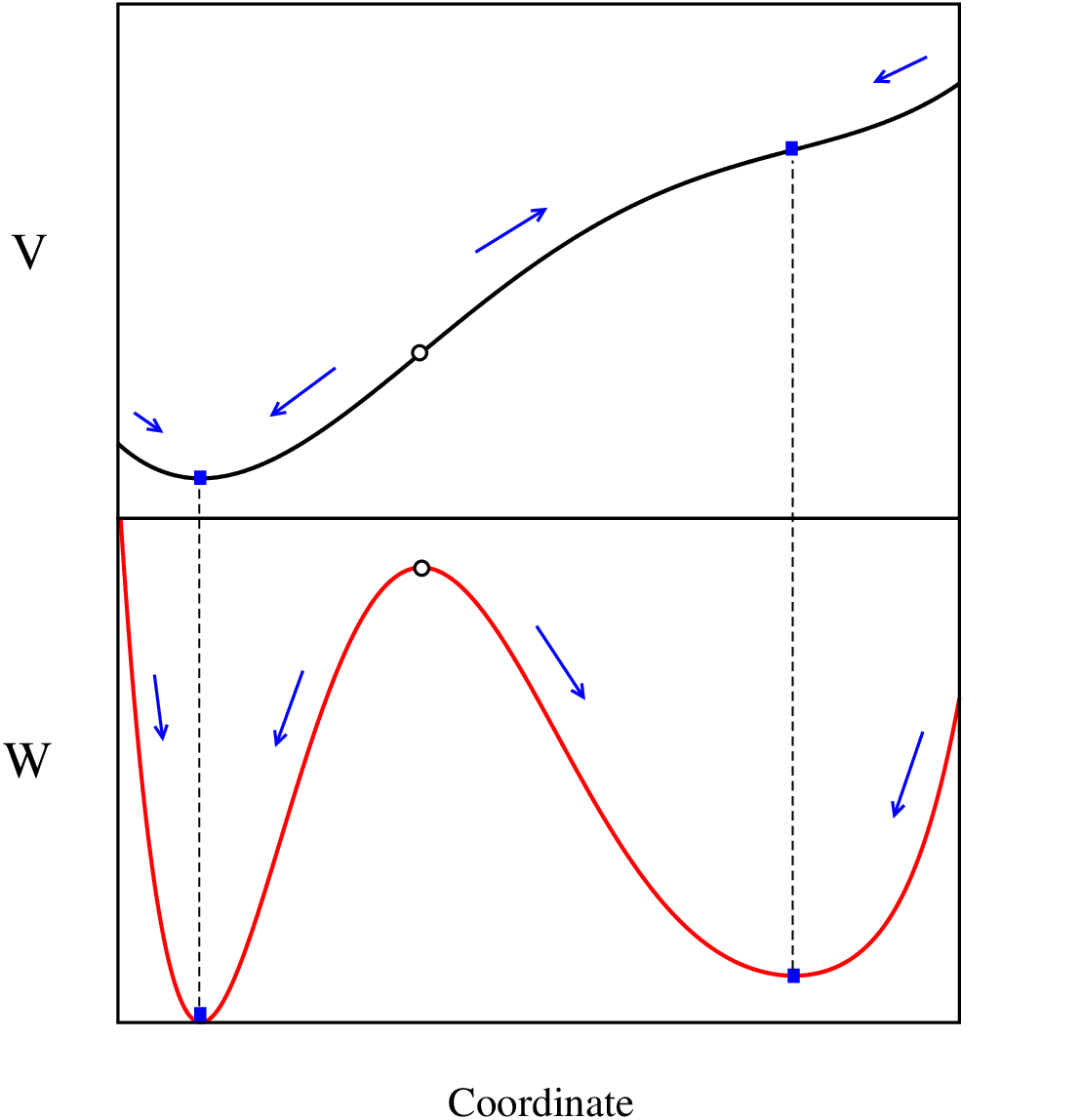}
\caption{Example of profile of $V$ and $W$ along a given direction. The inflection point of $V$ corresponds
to a local minimum of $W$ with $W>0$. The arrows indicate the basins of attraction of the minima of $W$.}
\label{fig_pro_2}
\end{figure}

The aim of our approach is to use a method of investigation of the PES that allows us to reject all 
the {\it non-relevant} degrees of freedom for the description of the slow supercooled dynamics.
The idea is that there are particular points of the PES associated in some way to the instantaneous 
configurations, and that the properties of these points contain all the important information 
for the determination of the slow dynamics. 
Possible and natural candidates are the saddle points of the PES.
In principle one can think to partition the whole configuration space in basins of attraction of saddles
(the definition of the partition is obviously not unique, a possible choice could be that obtained using 
the Voronoi polyhedra in the $3N$-dimensional space). 
However a useful partition in basins of saddles (that is local, i.e. allows us to associate 
a saddle point to each instantaneous configuration using only local information of the PES) is still not available.
A possible partition is that obtained using the pseudo-potential \cite{weber} $W=|\nabla V|^2$, 
associating to each instantaneous configuration during the molecular dynamics a minimum of $W$:
\begin{center}
{\it instantaneous configuration $\longrightarrow$ minimum of $W$} \ .
\end{center}
However, as already pointed out, the relation between all the minima of $W$ and saddles of $V$ is not one-to-one:
absolute minima of $W$ (with $W=0$) are all saddles of $V$
\begin{center}
{\it absolute minima of $W$ $\longleftrightarrow$ saddles of $V$} \ ,
\end{center}
but local minima (with $W > 0$) are inflection points of $V$   
(more precisely they are inflection points with the first derivative of $V$ of the same sign 
of the first derivative of the curvature in the inflection direction):
\begin{center}
{\it local minima of $W$ $\longrightarrow$ inflection points of $V$} \ .
\end{center}
In Fig.s \ref{fig_pro_1} and \ref{fig_pro_2} 
examples of two possible one-dimensional profiles of $V$ and corresponding $W$ are shown as 
a function of a generic one-dimensional coordinate. 
In Fig. \ref{fig_pro_1} the two absolute minima 
of $W$ correspond to two saddle points of $V$ (in one dimension a minimum and a maximum).
The arrows define the basins of attraction of minima of $W$, and the boundaries of these basins 
are indicated with open symbols.
In Fig. \ref{fig_pro_2} the local minimum of $W$ corresponds to an inflection point of $V$.

We note that at inflection points the eigenvalues of the Hessian vanish, so in principle 
one can recognize them simply analyzing the eigenvalues of the Hessian of $V$. 
In a previous work \cite{noi_selle} we used this eigenvalues criterion (choosing a threshold to discriminate 
inflection directions) and we found that local minima of $W$ were found very rarely.
A closer inspection, however, has revealed that, due to ``imperfect'' minimizations of $W$,
the eigenvalues associated to a point ``close'' (due to the not perfect minimization)
to an inflection point were of the same order of magnitude of 
the eigenvalues associated to a saddle.
Refining the analysis of the minima of $W$, we find that the number of points that are true saddles
is very small (about $1$\% of the total), so almost all the minima of $W$ found are inflection points.
However there is another interesting characteristic that has to be considered, 
the number of inflection directions (number of zero eigenvalues of Hessian, excluding the tranlations)
at the minima of $W$: 
if this number is small one can think of the minima of $W$ as a ``quasi'' saddle point QSP, 
as they are {\it true} saddle points in the subspace orthogonal to the small number of inflection directions.
One can conjecture that the properties of a QSP are very similar to that of a true saddle point 
associate to an instantaneous configuration; however, a relationship between them has still to be proved.
In the simulated MLJ system the number of inflection directions $n_0$ 
with respect to the number of negative curvatures $n_{_S}$
are reported in Table \ref{table1} for the different investigated temperatures.
The values of $n_0$ (from $1$ to $4$) indicate that they correspond to few directions in the configuration space;
a small value that allows us to speak about QSP as good candidates to
approximate the properties of the true saddle points.
We note that the fraction $f=n_0/n_{_S}$ is higher for the low temperature data (where the quantity $n_{_S}$
is small), evidencing how these points are more influenced by inflection directions. 
However, also in this case it is possible to estimate the error in the calculation of the order 
(see Fig. \ref{n_vs_T_2} in the next section), evidencing the robustness of the obtained results.

\begin{table}
\caption{\label{table1}Average number of negative curvatures ($n_{_S}$) and inflection directions ($n_0$),
and their ratio (f) at minima of $W$ for different temperatures.}
\begin{ruledtabular}
\begin{tabular}{cccc}
T 
&$n_{_S}$\footnote{Absolute number of negative curvatures} 
&$n_0$\footnote{Absolute number of zero curvatures (excluding translations)}
&f\footnote{Fraction of inflection directions $n_0/n_{_S}$} \\
\hline
0.49 & 1.4  &  1.1 &  0.78 \\
0.53 & 3.0  &  2.4 &  0.80 \\
0.57 & 5.6  &  2.5 &  0.45 \\
0.61 & 7.0  &  3.1 &  0.44 \\
0.64 & 7.6  &  2.8 &  0.37 \\
0.71 & 10.3 &  2.9 &  0.28 \\
0.81 & 15.9 &  3.6 &  0.23 \\
0.97 & 20.3 &  3.4 &  0.17 \\
1.13 & 24.5 &  4.3 &  0.18 \\
1.28 & 30.1 &  3.9 &  0.13 \\
1.44 & 32.9 &  3.9 &  0.12 \\
1.57 & 36.2 &  4.4 &  0.12 \\
\end{tabular}
\end{ruledtabular}
\end{table}

A natural question now arises: what are the relevant features of the found minima of $W$ that are useful to describe the 
long time dynamics of the supercooled liquids analyzed ?
In the spirit of our early simple conjecture that the relevant quantity is related to the number of negative
curvatures at saddle points, we speculate that 
the relevant information is in the negative curvatures and not in the few inflection directions.
So, no matter if they are {\it true} or {\it quasi} saddle points, we are going to check if their properties are
able to describe and to give a microscopic interpretation of the dynamic processes in the supercooled regime.

\section{Dynamic transition}

We firstly analyze the temperature behavior of some quantities related to quasi saddle points, 
and compare them to the corresponding behavior of the same quantities calculated at instantaneous configurations 
and at inherent structures.

For each temperature (ranging from $T \sim 1.6$ to $T\sim T_c$), we analyze $20$ independent equilibrium configurations.
For each configuration we calculate the associated IS, implementing a steepest descent algorithm which moves in the direction of 
$-{\bf \nabla}V = {\bf F}$,
and the associated minimum of $W$, moving in the direction of $-{\bf \nabla}W = {\cal H} \cdot  {\bf F}$, where 
${\cal H}$ is the Hessian matrix. 
So, at each temperature, we have three different points of the configuration space: 
the instantaneous configuration ${\bf r} $, 
the inherent structure ${\bf r}_{_{IS}}$, 
and the quasi saddle configuration ${\bf r}_{_S}$.

\begin{figure}[htb]
\vspace{.5cm}
\includegraphics[width=.45\textwidth,angle=0]{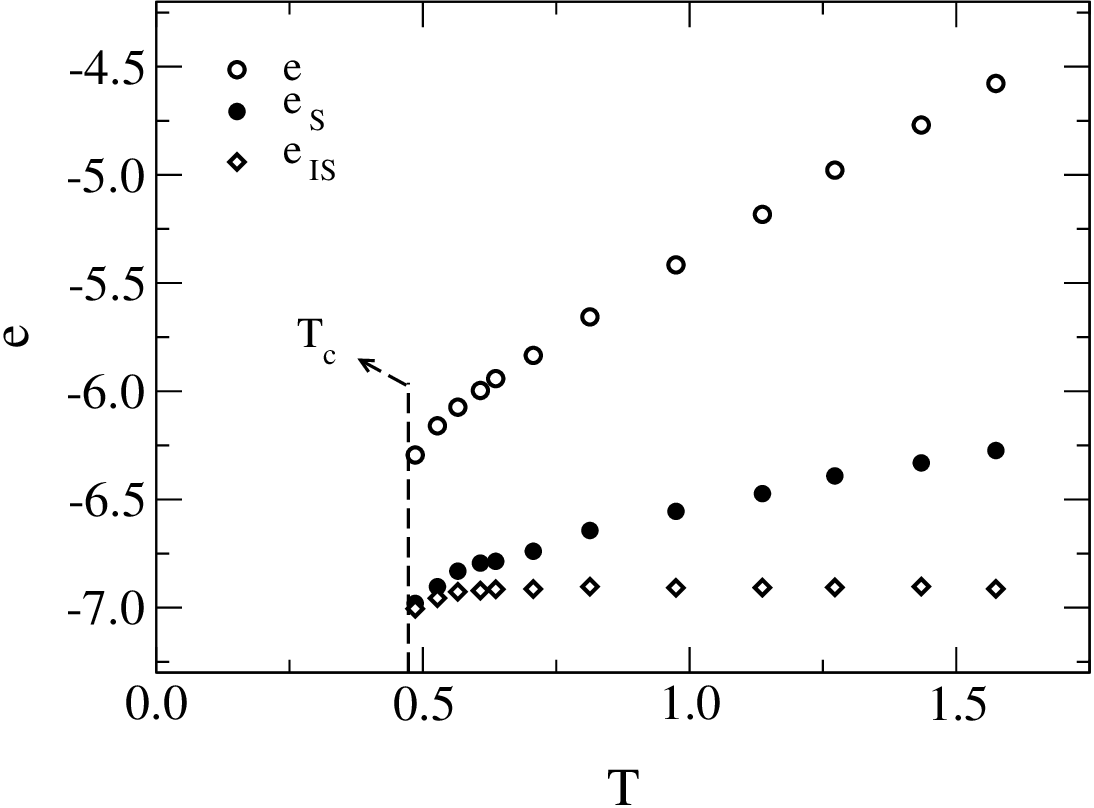}
\caption{Instantaneous $e$, saddle $e_{_S}$, and inherent structure $e_{_{IS}}$ energies as a function 
of temperature.}
\label{e_vs_T}
\end{figure}

\begin{figure}[b]
\vspace{.5cm}
\includegraphics[width=.45\textwidth,angle=0]{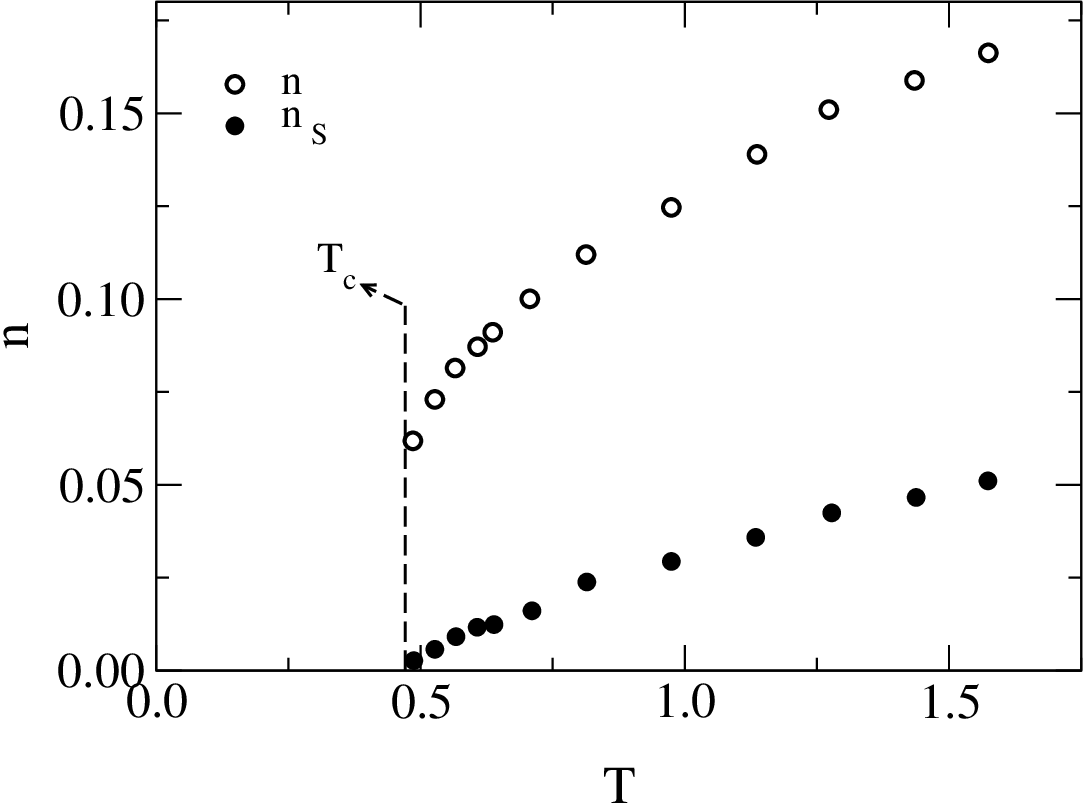}
\caption{Instantaneous order $n$, and saddle order $n_{_S}$  as a function of temperature $T$.}
\label{n_vs_T_1}
\end{figure}

\begin{figure}[htb]
\vspace{.2cm}
\includegraphics[width=.45\textwidth,angle=0]{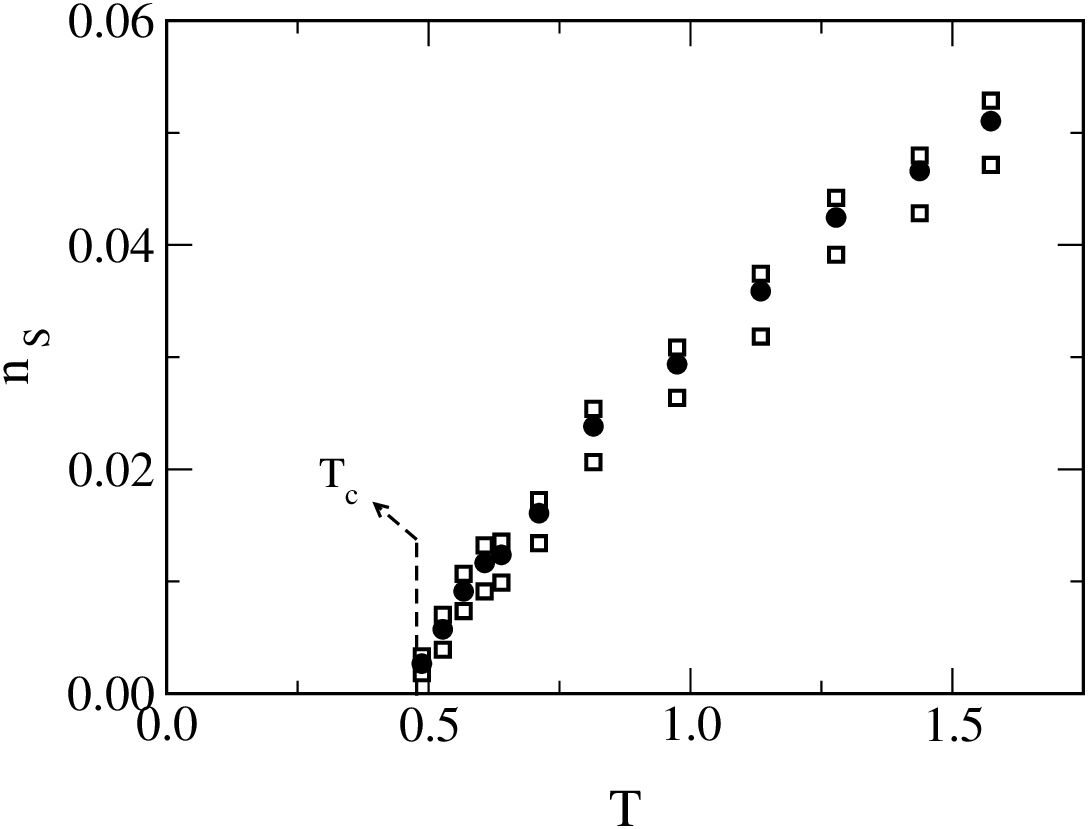}
\caption{Saddle order $n_{_S}$ as a function of temperature. The open squares are the upper and lower estimations
considering the ``false'' directions (see text for details).}
\label{n_vs_T_2}
\end{figure}

In Fig. \ref{e_vs_T} the temperature dependence of the energies of the different points are shown: $e $, $e_{_{IS}}$ and $e_{_S}$.
The value of the mode coupling temperature $T_c=0.475$ (obtained by diffusivity data, see section IV) is also indicated. 
The energy of the inherent structures shows a constant behavior down to $T\sim 0.8$, below which it starts to decrease abruptly
(note however the small variation range with respect to that of the $e_{_S}$).
Below this temperature the system starts to visit basins of minima of lower and lower energy.
The quantity $e_{_S}$ shows a different behavior, decreasing gradually with temperature and approaching $e_{_{IS}}$ close to $T_c$.
The fact that $e_{_S}$ lies well below $e $ indicates that the process of minimization of the pseudo-potential $W$
consists in a downward path in the PES along the majority of the degrees of freedom.
The coincidence of $e_{_{IS}}$ and $e_{_S}$ at $T_c$ suggests the hypothesis that the sampled saddles start 
to be mainly minima at this temperature.
The answer to this question  is in the behavior of the number of negative curvatures as a function of temperature.
In Fig. \ref{n_vs_T_1} the fraction of negative curvatures (number of negative curvatures over the total degrees of freedom $3N$)
is shown for the analyzed configurations ${\bf r} $ ($n $) and ${\bf r}_{_S}$ ($n_{_S}$)
(obviously this number is zero for the inherent structures ${\bf r}_{_{IS}}$).
The first thing to observe is that the quantity $n_{_S}$ lies well below the instantaneous order $n $, indicating that the 
process of minimization of $W$ leads to a point well below in the PES, with a number of negative curvatures
less than that at the equilibrium starting point.
We have then found a direct method to reject many negative eigenvalues (corresponding to not diffusive directions)
of the Hessian at the instantaneous configurations.
If the remaining negative curvature directions are those useful to describe the slow dynamics of the system, will be soon clear.
A first good indication of the relevance of the quantity $n_{_S}$ for the description of the dynamics is the fact that 
its well defined temperature behavior extrapolates to zero very close to the $T_c$ value. 
In Fig. \ref{n_vs_T_2} the temperature behavior of $n_{_S} (T)$ is shown in a larger scale.
It is shown the $n_{_S}$ calculated in a direct way (as the fraction number of negative curvatures at the minima of $W$)
and an upper and lower estimation of the ``true'' saddle order, considering all the inflection directions as if they 
contributed to the saddle order (upper points) 
and as if they didn't contribute to it (lower points).
Fitting the three different data with a power law $(T-T_0)^\gamma$,
we obtain very similar value of $T_0 \sim T_c$,
indicating the robustness of the analysis.
  
The sampled minima of $W$ are then QSP with a number of negative curvatures that decreases lowering the
temperature, and approaches zero at the temperature of the dynamic mode coupling transition.
At $T_c$ and below it, $T<T_c$, the sampled saddles are in average minima of $V$, i.e. saddles with order zero.
The emergent scenario for the dynamics is then the following:
\begin{itemize}
\item for $T>T_c$ the system lies with high probability close to borders of basins of attraction of inherent structures,
\item for $T<T_c$ the system spends most of the time trapped in the basins of inherent structures.
\end{itemize}
These findings confirm the conjecture that $T_c$ marks a dynamics crossover, 
evidencing a possible structural PES interpretation
of this change in the dynamics.

\begin{figure}[htb]
\vspace{.5cm}
\includegraphics[width=.45\textwidth,angle=0]{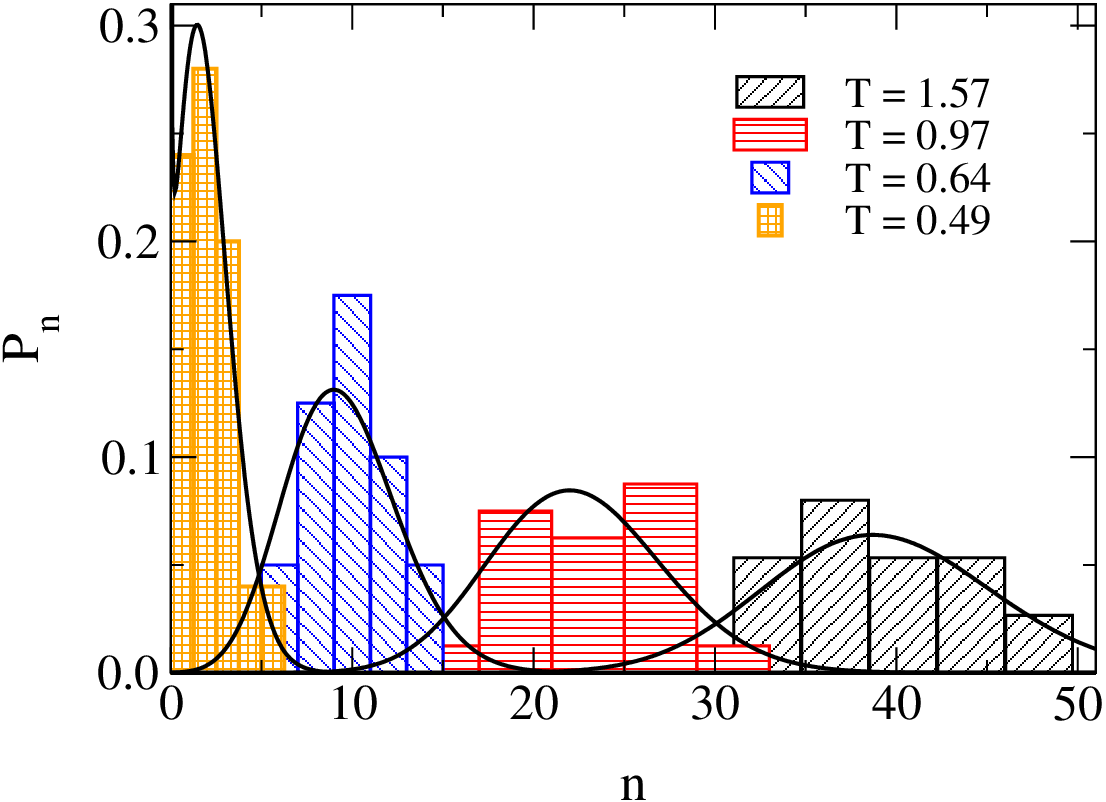}
\caption{Distribution of saddle order $n$ (the absolute number of negative curvatures) for four 
different temperatures. The line are the Poisson distributions (see the text).}
\label{distr}
\end{figure}

We conclude this section showing the distributions of the quasi saddle order for different temperatures.
In Fig. \ref{distr} the distribution $P_n$ of saddle order $n$ (now we indicate with $n$ the 
absolute saddle order, i.e. the total number of negative curvatures, not the fraction) is
shown for four different temperatures.
A simple conjecture allow us to obtain a prediction for these distributions.
Assuming the independence of the relevant slow degrees of freedom and 
a simple one dimensional profile of the independent coordinate (with only a minimum and a maximum) 
we have that the distribution of the saddle order $n$ of $N$ degrees of freedom is simply
the binomial distribution $P_n = {N \choose n} p^n (1-p)^{N-n}$, where $p$ is the probability 
that the single degree of freedom is close to (inside the basin of attraction of) the maximum.
In the large $N$ and small $p$ limit we obtain a Poisson distribution
\begin{equation}
P_n \sim \frac{<n>^n}{\Gamma (n+1)}\ e^{-<n>} \ ,
\label{pois}
\end{equation}
written in term of the $\Gamma$ function ($\Gamma (n+1)=n!$) and of 
the mean value of the saddle order $<n> = p \ N$ 
(the temperature dependence is in the $p$ parameter - small $p$ means low temperature).
Lines in Fig. \ref{distr} are the distributions obtained from Eq. \ref{pois}, 
with the $<n>$ value fixed by the data.
They well describe the measured distributions, suggesting the correctness of the simple viewpoint
of independent degrees of freedom 
(a simple model with this ingredient will be study in detail in the last section).

\section{Diffusion}

As demonstrated in the previous section, some features of the saddles probed during the 
dynamic evolution of the system are able to capture the dynamic crossover at $T_c$. 
In order to corroborate this finding we now go to analyze the diffusion properties in the supercooled regime
and to evidence a possible relationship with the saddle order $n_{_S}$.

\begin{figure}[htb]
\vspace{.2cm}
\includegraphics[width=.45\textwidth,angle=0]{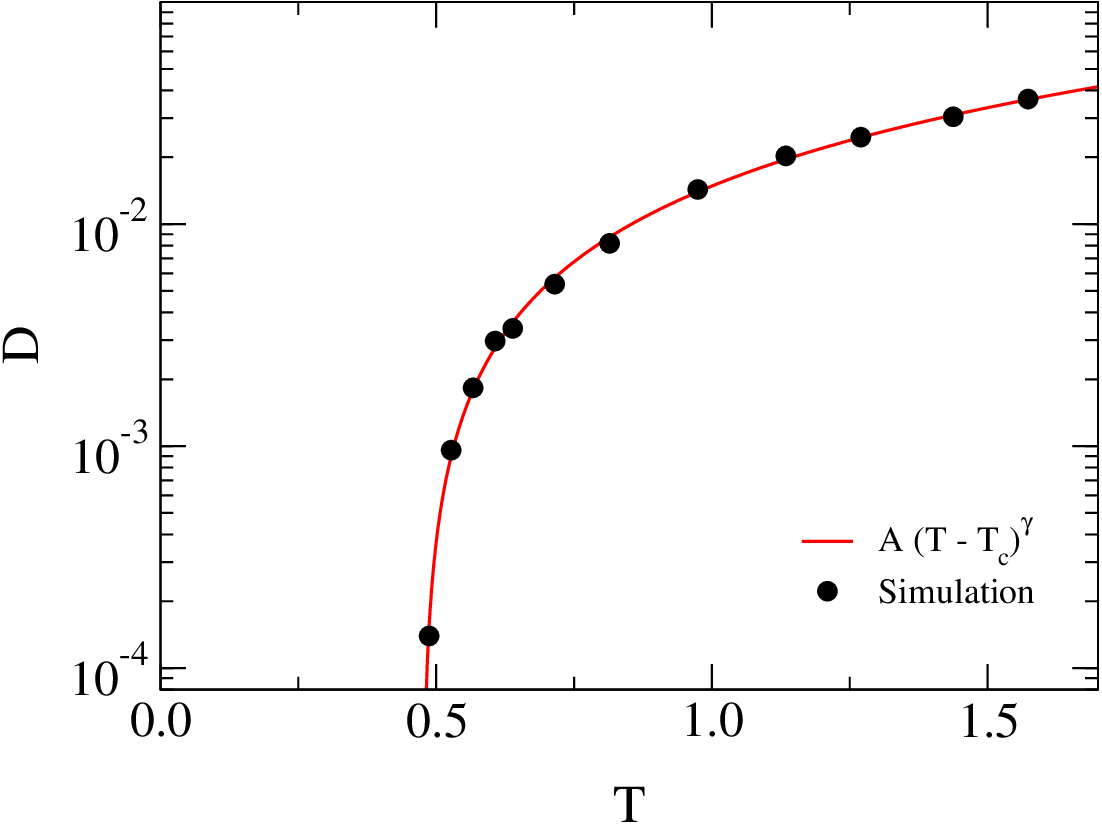}
\caption{Diffusion coefficient $D$ as a function of temperature.
The line is a power law fit with parameters reported in the text.}
\label{diffu}
\end{figure}

For different temperature above $T_c$ we have calculated the mean square displacement:
\begin{equation}
R^2 (t) = \frac{1}{N} \sum_i < |\vec{r}_i(t) - \vec{r}_i(0) |^2 > \ ,
\end{equation}
and, from its long time behavior, the diffusion coefficient $D$:
\begin{equation}
D = \lim_{t \rightarrow \infty} \frac{R^2(t)}{6t} \ .
\end{equation}
In Fig. \ref{diffu} the diffusion coefficient $D$ is shown as a function of temperature.
The mode coupling theory predicts a power law divergence of the inverse diffusivity close to $T_c$, 
so the behavior of the diffusion coefficient is a method to estimate $T_c$.
Fitting the data with power law (line in Fig. \ref{diffu})
\begin{equation}
D (T) \propto (T-T_c)^{\gamma} \ ,
\end{equation}
we obtain the value $T_c=0.475$.

The fact that also the saddle order $n_{_S}$ extrapolates to zero at $T_c$, suggests the possibility
of a relationship between $n_{_S}$ and the diffusivity $D$.
A direct prediction of a relation between the two quantities arises from a simple interpretation of the 
dynamic processes.
Supposing that the slow degrees of freedom responsible of the diffusion processes
are in some way related to the number of negative curvatures of the probed saddle points at a given temperature,
one can try to predict a relationship between the saddle order $n_{_S}$ and the diffusion coefficient.
Assuming that the number of diffusive directions at a given temperature are proportional 
to the saddle order at the same temperature, we can think of the diffusive process as a random walk process
in a space of dimension equal (or proportional) the the saddle order $n_{_S}$.
In this case one expects that the diffusion coefficient is proportional to the temperature times the saddle order:
\begin{equation}
D(T) \propto T \  n_{_S}(T) \ .
\label{d_vs_ns}
\end{equation}
We note that this hypothesis doesn't mean that in general all the negative curvatures at a 
generic saddle point are related
to a slow diffusive process, as there are saddles related to small local rearrangements of atoms.
It means only that the process of minimization of $W$, starting from an equilibrated configuration, leads in average
to a point (saddles or quasi-saddle doesn't matter) 
with negative curvature directions in some way 
related to the true diffusive directions.
\begin{figure}[htb]
\vspace{.2cm}
\includegraphics[width=.45\textwidth,angle=0]{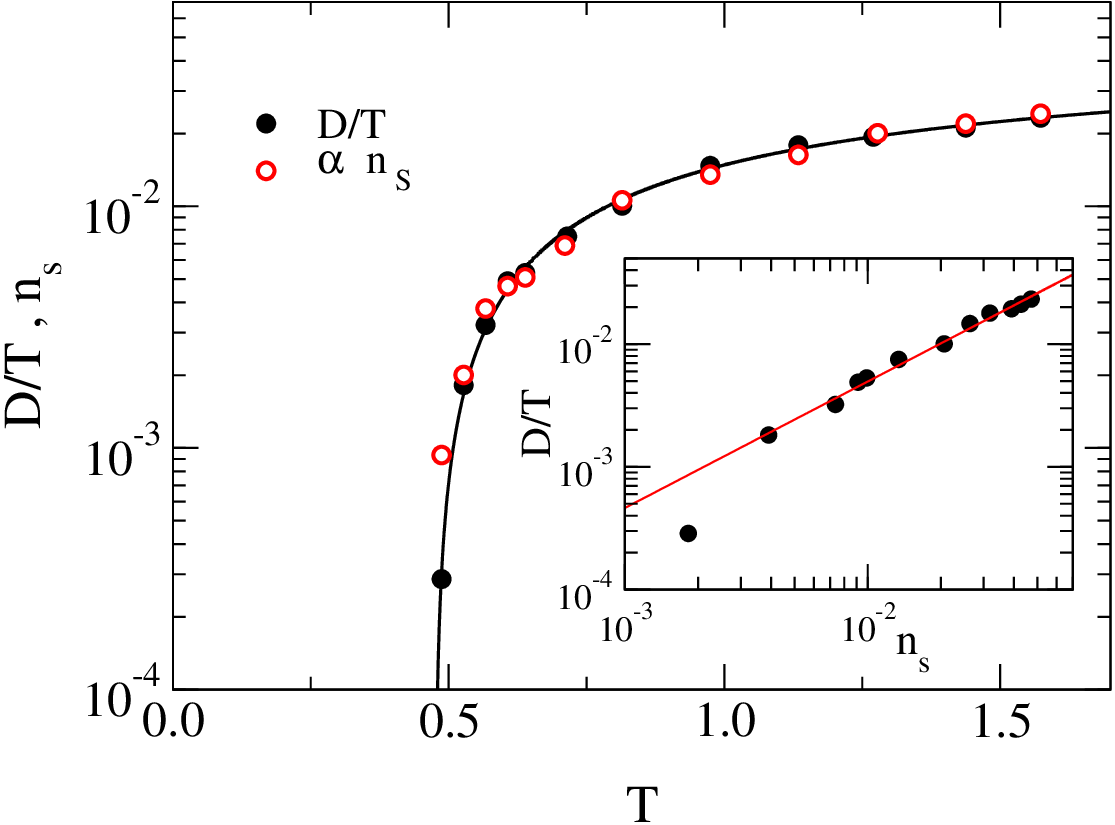}
\caption{Diffusivity over temperature $D/T$ and saddle order $n_{_S}$ (multiplied by a constant factor $\alpha$)
as a function of temperature.
The line is the power law fit of diffusivity.
In the inset the two quantities one against the other in a double log scale.
The line has slope equal to $1$.}
\label{ordiffu_2}
\end{figure}

In Fig. \ref{ordiffu_2} the diffusivity over the temperature calculated through molecular dynamics simulations
and the saddle order $n_{_S}$ are shown as a function of temperature: it seems that the simple random walk model on
a reduced space works well.
Eliminating the temperature and plotting the two quantities one against the other in a double log scale 
(inset of Fig. \ref{ordiffu_2}) we see that the Eq. \ref{d_vs_ns} is well suited,
except the last low temperature point,
which is however the most influenced by a possible not good equilibration procedure.
This kind of analysis deserves surely further investigations, for example analyzing in a more direct way the  
paths during the molecular dynamics evolution and their relation with the negative curvature directions at saddle points.
However one result emerges in a strong way: the quantity $n_{_S}$ seems to be very useful to reproduce 
dynamic processes and to give a possible PES interpretation of them.

\section{Landscape features}

\begin{figure}[htb]
\vspace{.8cm}
\includegraphics[width=.45\textwidth,angle=0]{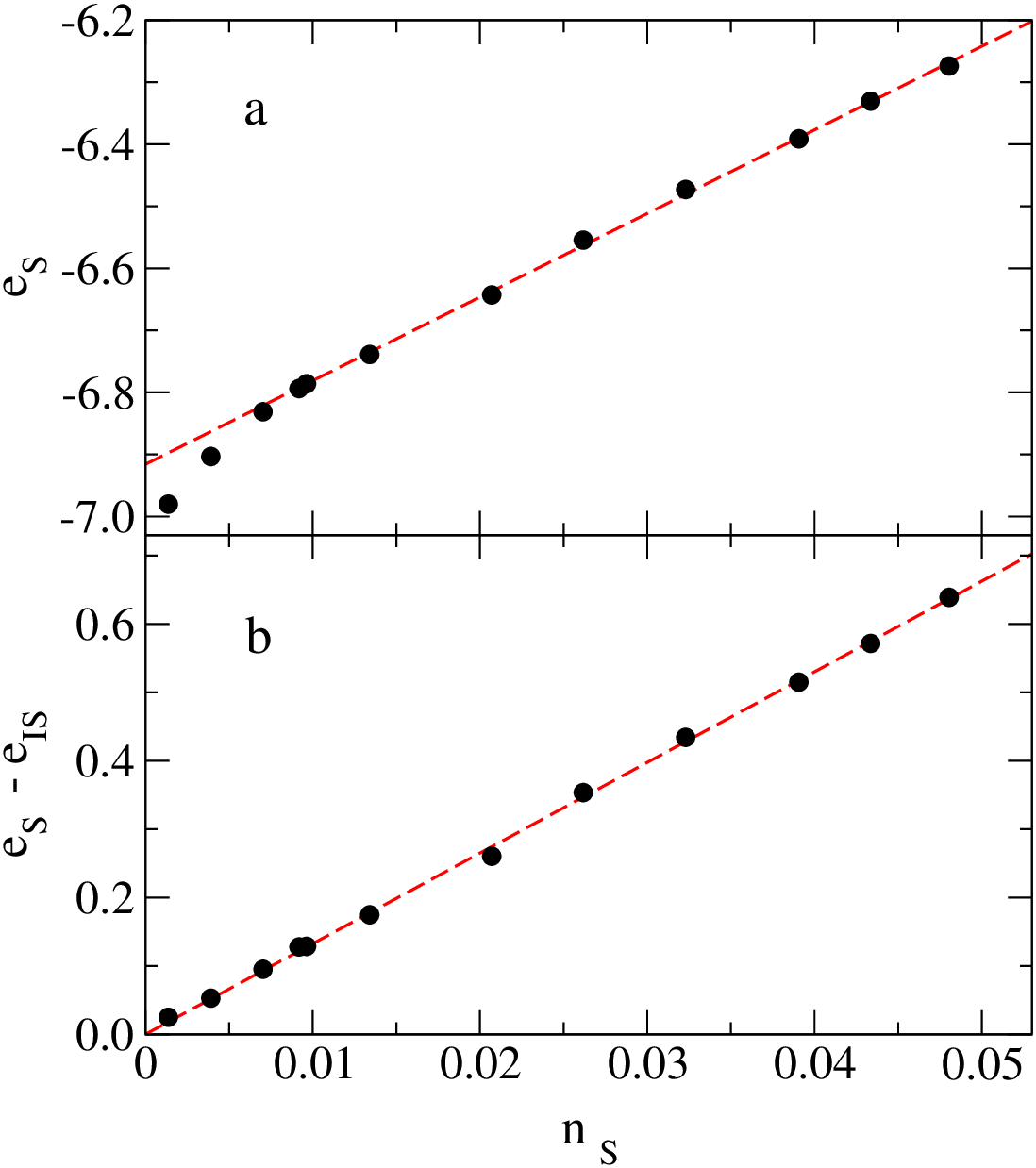}
\caption{ Saddle energy $e_{_S}$ (a) and saddle energy elevation from underlying minima $e_{_S} - e_{_{IS}}$ (b)
as a function of saddle order $n_{_S}$.}
\label{ene_ord_12}
\end{figure}

The previous analysis of the PES allows us to infer some topological characteristics of the landscape itself.
The first feature of the PES we analyze is the energy organization of the saddles, i.e. a possible relation
between energy and order of saddle points.
Plotting (see Fig. \ref{ene_ord_12}a)
the saddle energy $e_{_S}$ as a function of the saddle order $n_{_S}$ (using all the saddles found at different
temperatures), as done in \cite{cav_selle}, i.e.
\begin{equation}
e_{_S} (n_{_S}) \propto \Delta e' \ n_{_S} \ ,
\end{equation}
the data can be reproduced with $\Delta e' = 13.5$.
However, data at low energy (and order) deviate from the linear fit. 
Observing that these points are those sampled in the temperature range where the inherent structures
start to decrease in their energy values, one is attempted to relate the deviation from the linear fit of the 
energy-order relation to the inherent structure behavior.
In Fig. \ref{ene_ord_12}b the quantity plotted against the saddle order $n_{_S}$ is the elevation energy 
$e_{_S} - e_{_{IS}}$ of saddles
with respect to the corresponding local minima (minima visited at the same temperatures), obtaining now a remarkable
linear relationship in the full temperature range:
\begin{equation}
e_{_S} (n_{_S}) - e_{_{IS}} = \Delta e \  n_{_S} \ ,
\end{equation}
with $\Delta e = 13.3$.
This suggests that the energy landscape above a given minimum is organized in families of equally 
spaced energy saddle points, with only a single energy barrier parameter (in average) $\Delta e$,
that represents the energy gap between a saddle of order $n_{_S}$ and a saddle of order $n_{_S}+1$.
A possible explanation of this result is the following: 
saddles of a given order are obtained from a combination of independent saddles of order $1$, 
that lie a fixed amount over minima.
One can think of the excitations as as a ``gas'' of non-interacting degrees of freedom
(in the next section we develop this viewpoint in a deeper way, introducing a simple landscape model).
Similar results are found in the out-of-equilibrium regime
\cite{noi_aging}, suggesting that the PES properties obtained are not influenced by the kind of dynamics the
system use to explore its PES.

Another interesting topological information about the PES is obtained calculating the distance relations between 
adjacent inherent structures.
For each saddle point we have perturbed the system along a randomly chosen
negative curvature direction and then started 
a minimization procedure of the potential $V$ in order to find the underlying minimum. The same minimization
has been performed perturbing the system along the previous negative curvature direction but in the opposite versus.
In this way we obtain couples of inherent structures associated to each negative direction (we call them 
{\it adjacent} inherent structures).
We then evaluated the Euclidean distance between adjacent IS, defined as
\begin{equation}
d^2_{_{IS}} = \frac{1}{N} \sum_i < |\vec{r}_i (IS_1) - \vec{r}_i (IS_2) |^2 > \ ,
\end{equation}
where $IS_1$ and $IS_2$ are two adjacent minima and the average is over different couples of adjacent IS.
\begin{figure}[b]
\centering
\vspace{.3cm}
\includegraphics[width=.45\textwidth,angle=0]{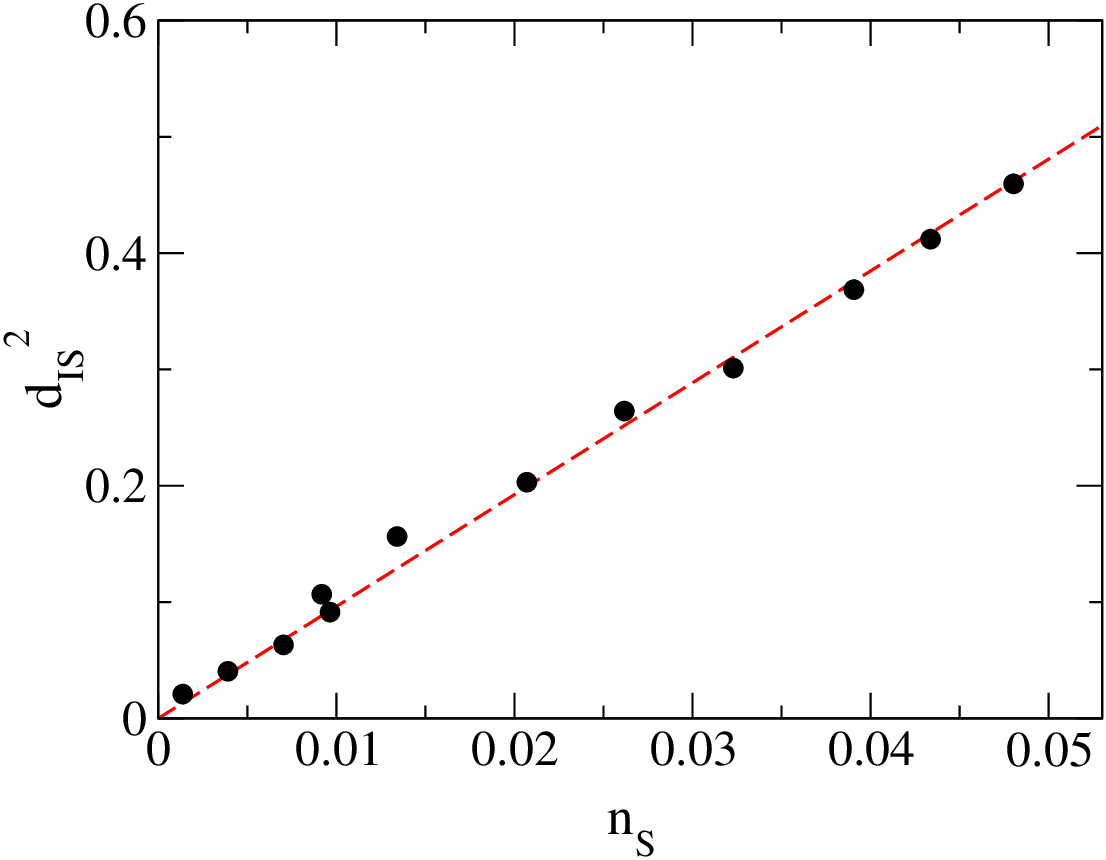}
\caption{The mean square distance between adjacent minima $d_{_{IS}}^2$ vs. saddle order $n_{_S}$.}
\label{d_order}
\end{figure}
In Fig. \ref{d_order} the quantity $d^2_{_{IS}}$ is shown as a function of the order $n_{_S}$ of the starting saddle.
The observed linear relation between the two quantities
\begin{equation}
d^2_{_{IS}} (n_{_S}) = d_0^2 \  n_{_S} \ ,
\end{equation}
(with $d_0^2=9.6$) suggests a simple topological interpretation:
the descent path from a saddle of order $n_{_S}$ towards the underlying minima can be view as a sequence of independent 
random steps, each of them decreasing the saddle order by $1$ and increasing the mean square distance 
between the underlying local minima by a constant amount $d_0^2$. 

The above results seem to indicate that in some aspects the PES exhibits a 
very simple and organized structure.
This suggests the possible use of simple solvable mathematical models 
in order to capture some of the relevant features of the PES 
explored in the supercooled regime.

\section{Trigonometric Model}
A very simplified model that is able to capture some of the regular characteristics of the PES of liquids 
evidenced above, is the so called Trigonometric Model (TM), introduced by Madan and Keyes \cite{madan_keyes}.
In this section we calculate for the TM the behavior of the quantities 
previously analyzed for the Lennard-Jones system.

The TM is a model for $N$ independent degrees of freedom with Hamiltonian:
\begin{equation}
H_{_{TM}} = \Delta  \sum_i [1 - \cos ( \varphi_i )] \ ,
\label{trigm}
\end{equation}
where $\{\varphi_i\}$ are angular variables: $\varphi_i \in [0,2\pi)$.
The PES of TM reproduce the regularity of the average saddle properties of the Lennard-Jones PES:
saddles of order $n_{_S}$ are $2\Delta$ over (in energy) and $\pi$ near (in distance) saddles of order $n_{_S}-1$.
The thermodynamics is easy computable, as the partition function is factorized:
\begin{equation}
Z_{_{TM}} (\beta) = Z_0^N(\beta) \ ,
\end{equation}
where $\beta=T^{-1}$ (we use the unit $K_B=1$) and 
\begin{equation}
Z_0(\beta) = \int_0^{2\pi} d\varphi \ e^{-\beta \Delta [1 - cos ( \varphi )]} =
2 \pi\ e^{-\beta \Delta} \  I_0(\beta \Delta) \ ,
\end{equation}
where $I_0$ is the Bessel function of order zero.
The energy density $e_{_{TM}}= - N^{-1} \partial_{\beta} \log ( Z_{_{TM}} )$ is easy written in term of $I_0$
and $I_1$ (the Bessel function of order $1$, $I_1(x)=I_0'(x)$):
\begin{equation}
e_{_{TM}} (\beta) = \Delta \left[ 1 - \frac{I_1(\beta \Delta)}{I_0(\beta \Delta)} \right] \ .
\end{equation}
\begin{figure}[htb]
\centering
\vspace{.2cm}
\includegraphics[width=.45\textwidth,angle=0]{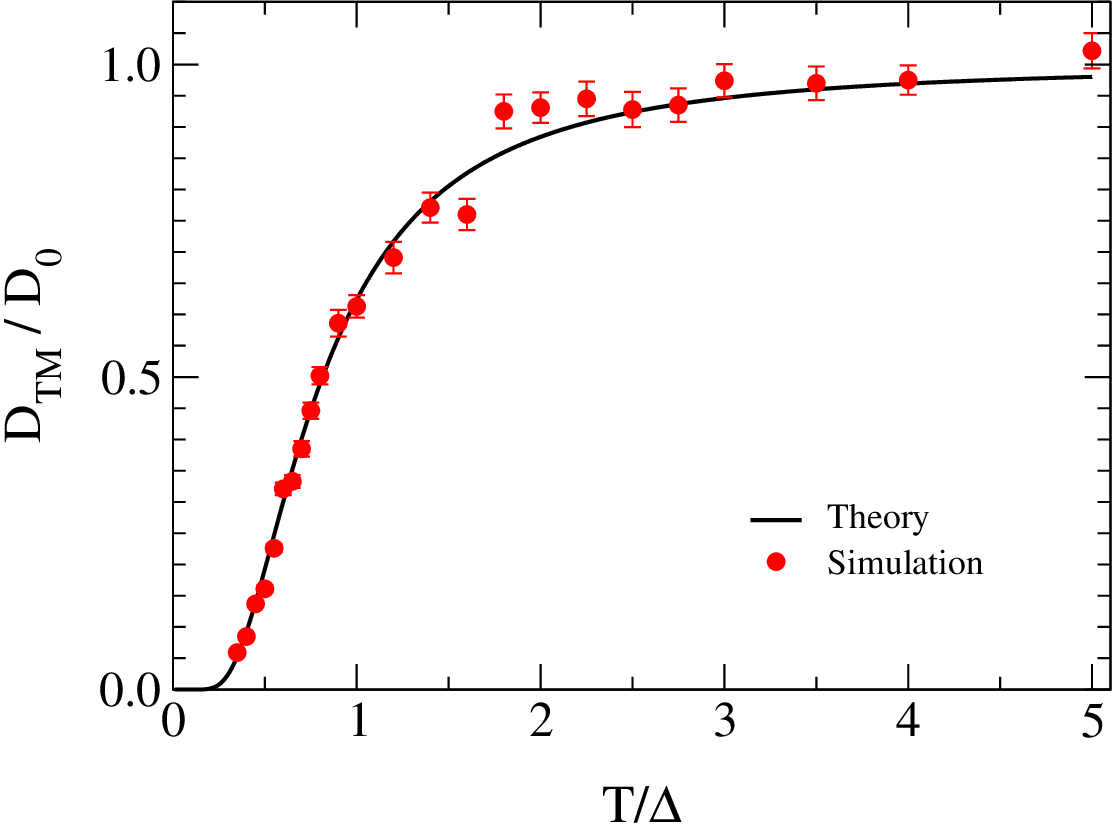}
\caption{The reduced diffusivity $D_{_{TM}}/D_0$ as a function of reduced temperature $T/\Delta$.
The symbols are from iso-potential simulation (see the text).}
\label{fig_diffuTM}
\end{figure}

In order to test the reliability of TM in the description of liquid behavior, we now calculate the 
diffusion coefficient and the saddle order as a function of temperature.
The diffusion coefficient $D_{_{TM}}$ is exactly computable calculating the mobility in linear response
\cite{risken}, with the introduction of a  Langevin dynamics:
\begin{equation}
\gamma \dot{\varphi_i} (t) = F - \nabla_i H_{_{TM}} + \eta (t) \ , 
\end{equation}
where $F$ is an external force, $\gamma$ is the friction, 
$\eta$ is a random variable with mean zero and $\delta$-correlate in time:
$<\eta(t) \eta(t')> = (2 \gamma / \beta) \delta (t-t')$ (we choose $m=1$ mass unit).
Following \cite{risken} we obtain:
\begin{equation}
D_{_{TM}} = \frac{1}{\gamma \beta \ I_0^2 (\Delta \beta)} \ .
\end{equation}
Using the expansion of $I_0$ for large argument 
( $I_0 (x) \sim \exp(x) / \sqrt x$, for large $x$) 
we obtain the Arrhenius behavior for the diffusivity at low temperature:
\begin{equation}
D_{_{TM}} \propto e^{-2\beta \Delta} \hspace{.5cm} {\rm for\  low\ } T \ 
\label{arre}
\end{equation}
(we note the value $2 \Delta$ is the energy barrier value for the single degree of freedom).
The dimensionless quantity 
\begin{equation}
\frac{D_{_{TM}}}{D_0} = I_0^{-2} (\Delta \beta) \ ,
\end{equation}
where $D_0=(\gamma \beta)^{-1}$ is the free Brownian diffusion coefficient,
is shown in Fig. \ref{fig_diffuTM} as a function of temperature (line).
\begin{figure}[htb]
\centering
\vspace{.2cm}
\includegraphics[width=.47\textwidth,angle=0]{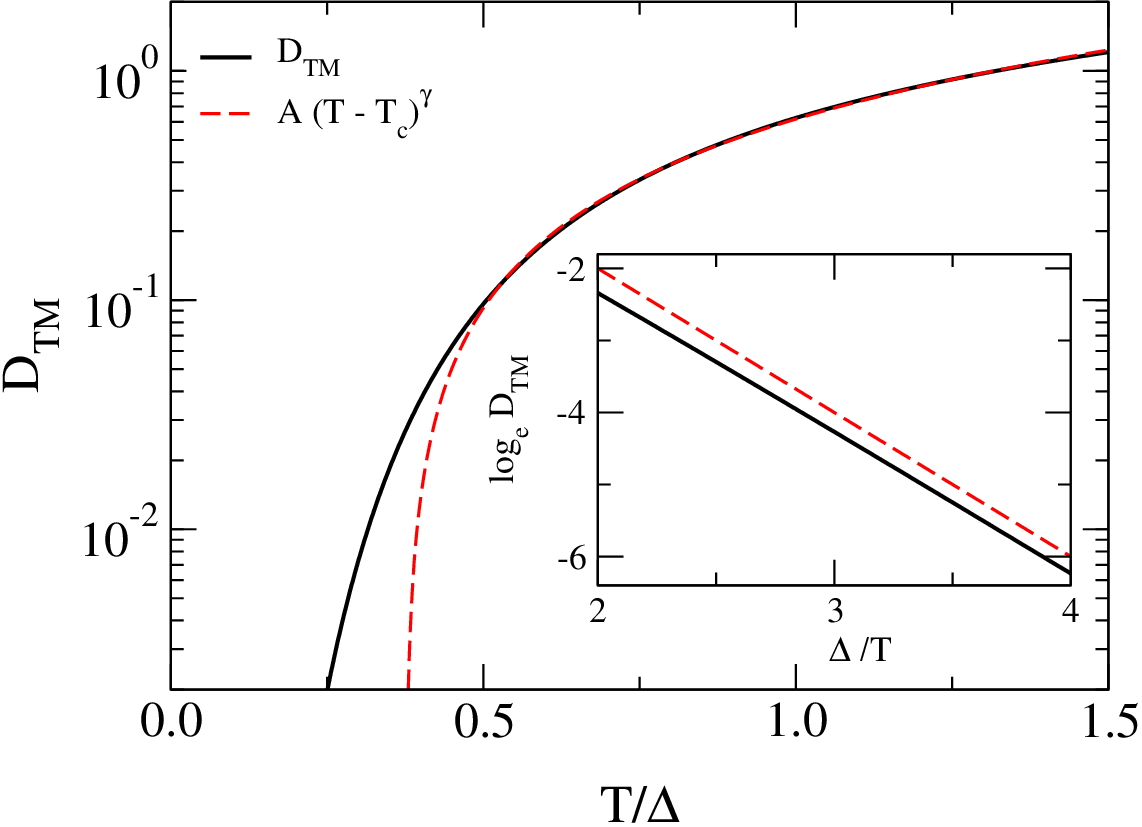}
\caption{The diffusivity $D_{_{TM}}$ as a function of reduced temperature $T/\Delta$ in a logarithmic plot (full line).
The dashed line is the power law fit of the high temperature range ($T_c/\Delta = 0.38$, $\gamma = 1.2$).
In the inset the low temperature behavior of diffusivity $D_{_{TM}}$ is shown as a function of inverse temperature
in an Arrhenius plot
(dashed line indicates slope $2$).}
\label{fig_diffu1TM}
\end{figure}
An interesting observation arises from the comparison between this theoretical canonical result
and a numerical iso-potential (dynamics at constant potential energy) calculation.
The points in Fig. \ref{fig_diffuTM} are obtained from a simulation of $N=1000$ variables constrained 
to move at constant potential energy (the dynamics is a random walk dynamics in which the forces 
$- \nabla_i H$ are projected to the constant potential energy surface). 
The coincidence between analytical and numerical data evidences 
how canonical and iso-potential calculations lead to the same equilibrium dynamic properties.
In Fig. \ref{fig_diffu1TM} the diffusion coefficient $D_{_{TM}}$ is shown as a function of temperature
in a semi-logarithmic scale. 
We note that, though $D_{_{TM}}$ is a smooth function, one can define a change in the behavior extrapolating 
a critical temperature $T_c$, using a power law fit in the low temperature range 
(from $0.5$ to $1.5$):
\begin{equation}
D_{_{TM}} (T) \propto (T-T_c)^{\gamma} \ ,
\end{equation}
obtaining the values: $T_c/\Delta = 0.38$ and $\gamma = 1.2$.
Below $T_c$ the diffusion coefficient can be approximated 
by  Eq. \ref{arre}, a low T Arrhenius behavior with 
energy barrier value $2 \Delta$ (see the inset in Fig. \ref{fig_diffu1TM}).

\begin{figure}[htb]
\vspace{.2cm}
\includegraphics[width=.45\textwidth,angle=0]{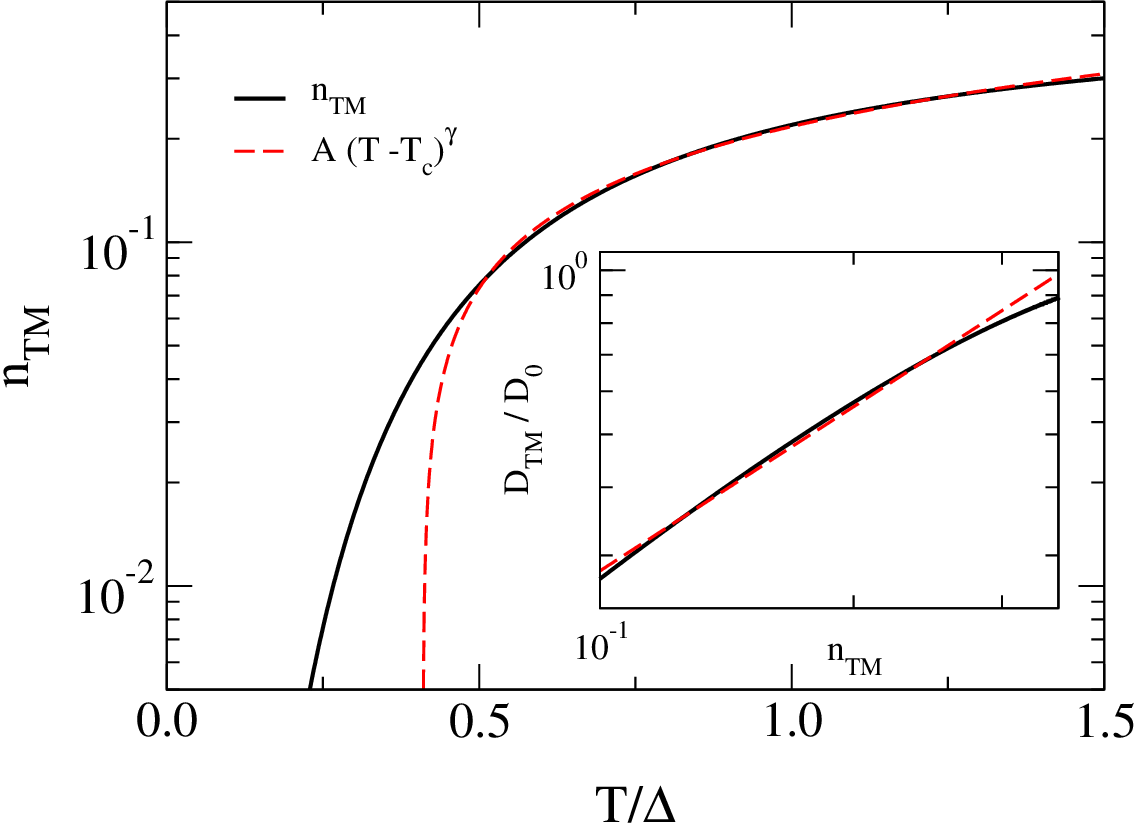}
\caption{ The saddle order $n_{_{TM}}$ as a function of reduced temperature $T/\Delta$.
The dashed line is the power law fit in the temperature range $0.5-1.5$ ($T_c/\Delta = 0.41$, $\gamma = 0.58$).
In the inset reduced diffusivity $D_{_{TM}}/D_0$ against the saddle order $n_{_{TM}}$ in a double logarithmic scale.
The dashed line has slope equal to $1$.}
\label{fig_saddTM}
\end{figure}

The saddle order $n_{_{TM}}$ is also computable in an exact way. 
Defining it as the negative curvatures at minima of $W = |\nabla H|^2$ (we note that, in the TM, all the minima
of $W$ are {\it true} saddles of $H$ and, moreover, the instantaneous order coincides with the saddle order) 
we can calculate the equilibrium mean value of $n_{_{TM}}$ as the 
probability that the variable $\varphi$ is in the range $[\pi/2,3\pi/2]$, where the $H$ profile as a negative curvature:
\begin{eqnarray}
\nonumber
n_{_{TM}} &=& Z_0^{-1} \int_{\pi/2}^{3\pi/2} d\varphi\ e^{-\beta \Delta [ 1 - cos (\varphi) ]} \\
&=&[2\pi I_0 (\beta \Delta) ]^{-1} \int_{\pi/2}^{3\pi/2} d\varphi\ e^{\beta \Delta  cos (\varphi) }
\ .
\end{eqnarray}
The temperature behavior of $n_{_{TM}}$ is shown in Fig. \ref{fig_saddTM}.
Also in this case it is possible to fit the data with a power law fit close to $T_c$
(from $0.5$ to  $1.5$),
obtaining a value of $T_c/\Delta = 0.41$,
close to that obtained from diffusivity ($T_c/\Delta=0.38$).

Having calculated the temperature behavior of diffusivity $D_{_{TM}}(\beta)$ and saddle order $n_{_{TM}}(\beta)$,
we can now try to study the relationship between them.
Due to the fact that a direct analytical expression of $D_{_{TM}}$ as a function of $n_{_{TM}}$ is not available,
we study it numerically eliminating the temperature parameter.
The relationship is shown in the inset of Fig. \ref{fig_saddTM},
where the quantity $D_{_{TM}}/D_0$ is plotted against $n_{_{TM}}$
in a logarithmic scale (this plot corresponds to that of the Lennard-Jones system 
- inset of Fig. \ref{ordiffu_2} - as the quantity 
$D_{_{TM}}/D_0$ is proportional to $\beta D_{_{TM}}$). 
The investigated $T$ range is close to the critical temperature $T_c$ (as in the liquid system) and 
also in this case we find an approximatively linear relationship (the dashed line in the inset of  Fig. \ref{fig_saddTM}
corresponds to a power law with exponent $1$):
\begin{equation}
D_{_{TM}} \propto T \  n_{_{TM}} \ .
\end{equation}
In conclusion, the very simple TM is able to reproduce some of the properties emerged in the analysis of the Lennard-Jones
liquid system, indicating that the use of simplified analytical  model of the PES 
is a promising field of research.

\section{Conclusions}

The analysis of the PES has been very useful in the investigation of supercooled 
dynamics of simple model liquids. 
Besides the two main approaches in the investigation of the PES, 
the instantaneous normal mode and inherent structures approaches,
we have discussed an additional sampling of the PES explored during the dynamic evolution of the 
system in configuration space: it consists in the analysis of the 
minima of the square gradient of the potential energy, $W= |\nabla V|^2$.
The characteristics of these sampled points seem to have a very close relationship with 
the transport properties of the system. 
The main result we obtained is the characterization of the dynamic transition temperature $T_c$
as the temperature at which the behavior of quantities related to the minima of $W$ 
changes on character: more specifically the number of negative eigenvalues of the Hessian
at these points extrapolates to zero at $T_c$. 
The minima of $W$ sampled at different temperatures appear to 
contain the relevant information about the diffusive directions and the process 
of minimization of $W$ eliminates the non-diffusive negative directions
usually present at instantaneous configurations.
Recently, an interesting technique to evaluate the non-diffusive directions
as been adopted in \cite{keyes_cgf}.
In some previous works we called these points {\it saddles}, as the absolute minima 
of the function $W$ are true saddle points of $V$.
However, following Doye and Wales \cite{doye_wales}
 a closer inspection has revealed that very often the sampled minima during 
dynamic evolution are local minima of $W$, that correspond to {\it inflection} points of $V$ 
(the energy profile along some normal mode directions at these points is an inflection direction).
The fact that the number of inflection directions at a given local minimum of $W$
is small, 
allows us to think that the main 
properties of these points are determined by the saddle directions (that give the value of the 
order, i.e. the number of negative curvatures). 
A further evidence of the relevance of the minima of $W$ is obtained analyzing the diffusivity.
The temperature behavior of the number of negative curvatures is related to that of the diffusion 
coefficient, indicating a close relationship between them.
It could be that the main information about the dynamic 
processes is in the saddle points and the relevant properties of the local minima of $W$ are 
good approximations of the properties of these {\it true} saddle points. 
But there is also the possibility that the {\it true} saddles are not
so relevant in the description of the dynamics, at least not more relevant than
points with inflection directions (local minima of $W$). 

All these findings lead to the following viewpoint about the supercooled dynamics:
the relevant diffusion processes happen in a subspace of the total configuration space
with dimension proportional to the number of saddle directions at the sampled minima of $W$,
and, due to the presence of these ``open'' directions, the relevant 
diffusive processes are not activated in energy.
Below $T_c$ things are different, as the number of negative directions at sampled minima
of $W$ is vanishingly small, indicating that the system is with high probability in a basin
of an inherent structure (minimum of $V$). 
In order to change basin and diffuse, now the system has to find the 
``good'' directions leading to other minima (an {\it entropic} process, considering the 
whole $3N$ dimensional space of the degrees of freedom, that could correspond
to an {\it energy activated} process in a suitable $n$ dimensional subspace - with $n \ll 3N$).

Another important result obtained refers
to  the structural organization of the PES, 
i.e. the relationship among the relevant points of it.
A very simple structure of the PES emerges:
saddles of order $n+1$ lie above saddles of order $n$ by a constant quantity $\Delta e$
and adjacent inherent structures are equally spaced in configuration space.
The above observed regularities suggest the possible use of simplified models of the PES
in order to capture relevant characteristics of the supercooled dynamics of liquids.
The model we analyzed here is the Trigonometric Model, introduced by Madan and Keyes \cite{madan_keyes},
a model of independent sinusoidal degrees of freedom. 
The behavior of the calculated diffusivity and saddle order seems to indicate
that the model, despite the absence of cooperativity, is able to reproduce
important aspects of liquid dynamics close to $T_c$. 
The study of more complex and realistic PES models, with interactions among the degrees of freedom,
is a interesting development of this kind of analysis.

\begin{acknowledgments}
We acknowledge financial support from INFM - {\it Iniziativa Calcolo Parallelo}.
We thank S. Ciuchi, C. De Michele and D.J. Wales for usefull discussions.
\end{acknowledgments}



\begin{thebibliography}{10}
\bibitem{nota}
The reason for which we still use the word ``saddle'' deserves further clarifications.
In our numerical experiments, although a few number of minima of $W$ are ``true'' saddles in 
a mathematical sense, most of the QSP found are ``almost'' saddles, in the sense that $W$
is almost zero numerically, and the number of inflection directions is very small.
\end{thebibliography}
\end{document}